\documentclass[a4paper]{jpconf}
\usepackage{graphicx}

\begin{document}
\title{No correlation between the Transit-Depth Metallicity of {\it Kepler} gas giant confirmed and candidates planets: A Bayesian Approach}


\author{Cyrine Nehm\'e$^{1,2}$ and Paula Sarkis$^3$}

\address{$^1$Department of Physics \& Astronomy, Notre Dame University- Louaize, PO Box 72, Zouk Mikael, Lebanon}
\address{$^2$CEA/DRF/Irfu/SAp, F-91191 Gif-sur-Yvette, France}
\address{$^3$Universit\"{a}t Bern Space Research \& Planetary Sciences Division Bern, Switzerland}
\ead{cnehme@ndu.edu.lb}

\begin{abstract}
Previous attempts to study the correlation between the transit depth and the stellar metallicity of {\it Kepler} gas giant planets has led to different results. A weakly significant negative correlation was reported from the  {\it Kepler's} (Q1-Q12) gas giant candidates with estimated radii of 5-20R$_\odot$ and [Fe/H] values taken from the Kepler Input Catalog (KIC). With the release of the last {\it Kepler} catalog (Q1-Q17), we now have the largest homogeneous sample of exoplanets. This enables a solid statistical analysis of this correlation. In the present work, we revise this correlation, within a Bayesian framework, for two large homogeneous samples: confirmed and complete. We expand a Hierarchical model to account for false positives in the studied samples. Our statistical analysis reveals no correlation between the transit depth and the stellar metallicity. The fact that we found no evidence of such correlation will have implications for planet formation theory and interior structure of giant planets. 
 
\end{abstract}

\section{Introduction}
NASA's Kepler Mission revolutionized the field of extra solar planets and now more then ever, it is possible to put statistical constraints on the observed planet properties and on theories of planet formation. Clues on the nature of giant planet formation might be revealed from the two correlations with stellar metallicity of main sequence stars hosting these planets. The first one is the correlation of the frequency of giant planets with stellar metallicity revealed by radial velocity surveys(\cite {FV}) and by transit surveys (\cite {WF}). The second correlation shows a positive trend between the mass of heavy-elements in giant planets and the stellar metallicity (\cite {MF}). This explosion of new information demonstrates the need to understand planet formation in general and presents an opportunity to compare observed trends to the theories of planet formation and evolution.
Earlier attempts to study the correlation between the transit depth and the stellar metallicity of {\it Kepler's} gas giant candidates has led to different results. For instance, \cite{dodo} reported a negative correlation with a weak significant value. The author studied the transit depth of 218 giant planets from \cite{Borukapj}, (Q1-Q12) catalog with estimated radii of 5-20R$_\oplus$ and the values of [Fe/H] taken from the {\it Kepler} Input Catalog (KIC). \cite{dodo} interpreted the negative correlation as evidence that metal rich planets of a given mass are denser than their metal poor counterparts leading to small radii (\cite{Fortnett}). Here, we will use the latest available catalog Q1-Q17 (\cite {MU}) to study correlation between the transit depth and stellar metallicity. Noting that a sample of stars with transiting planets may not accurately represent the true intrinsic distribution of the discovered planets. \cite{GM} reported the importance of including these effect since they can lead to biases in the properties of transiting planets and their host stars. For these reasons, we study only a subset of the target stars and the detected giant planets. We develop a flexible framework to account for uncertainties by expanding the Hierarchical Bayesian Model introduced by \cite {Kell}.

\section{Selection criteria and complete sub-sample }

We use the cumulative catalog of planets detected by the NASA {\it Kepler} mission which, as of April 2015, consisted of the latest Q1-Q17 catalog (\cite {MU}). Following \cite {dodo} and \cite {SL} we define gas giant planets as planets having a radius between 5-20R$_\oplus$. The stellar parameters were taken from the {\it Kepler} stellar Q1-Q16 database (\cite {HUB}). We ended up by having 84 planets confirmed and 305 candidates.\\
With the goal of performing a robust statistical method, we prepared two different samples. The first sample consists of all the 84 confirmed giant planets within the latest catalog. The second sample contains a complete subsample of both confirmed and candidates.  Hence, we performed cuts needed to take into account: the incompleteness of the catalog (\cite {Petig}), the selection effects for host stars and planets candidates, the detection efficiency and the false positives. After performing all the cuts (table \ref{cuts}), we retain 105 confirmed + candidates planets. We believe that thi complete subsample better represents the true intrinsic distribution of  {\it Kepler's} giant planets.

\begin{center}
\begin{table}[h]
\caption{\label{cuts}Summary of the cuts performed to obtain a complete subsample}
\footnotesize\rm
\centering
\begin{tabular}{@{}*{7}{l}}
\br
Parameter&Value\\
\mr
\verb"Stellar effective temperature", T$_{eff}$& 4000 - 7000 K\\
\verb"Stellar gravity", log {\it {g}} (cm/s$^{-2}$)& 4.0 - 5.0 \\
\verb"Stellar Radius", R$_\star$ & 0.7 - 1.4 R$_\odot$\\
\verb"Planetary Radius", R$_p$& 5 - 20 R$_\oplus$\\
\verb"Orbital Period", P& $<$ 90 days\\
\verb"Detection Efficiency", SNR& $>$ 18.8\\
\verb"Kepler magnitude", K$_p$& $<$ 16 mag\\
\br
\end{tabular}
\end{table}
\end{center}





\section{Method}

\begin{figure}
\centering
  \includegraphics[scale=0.7]{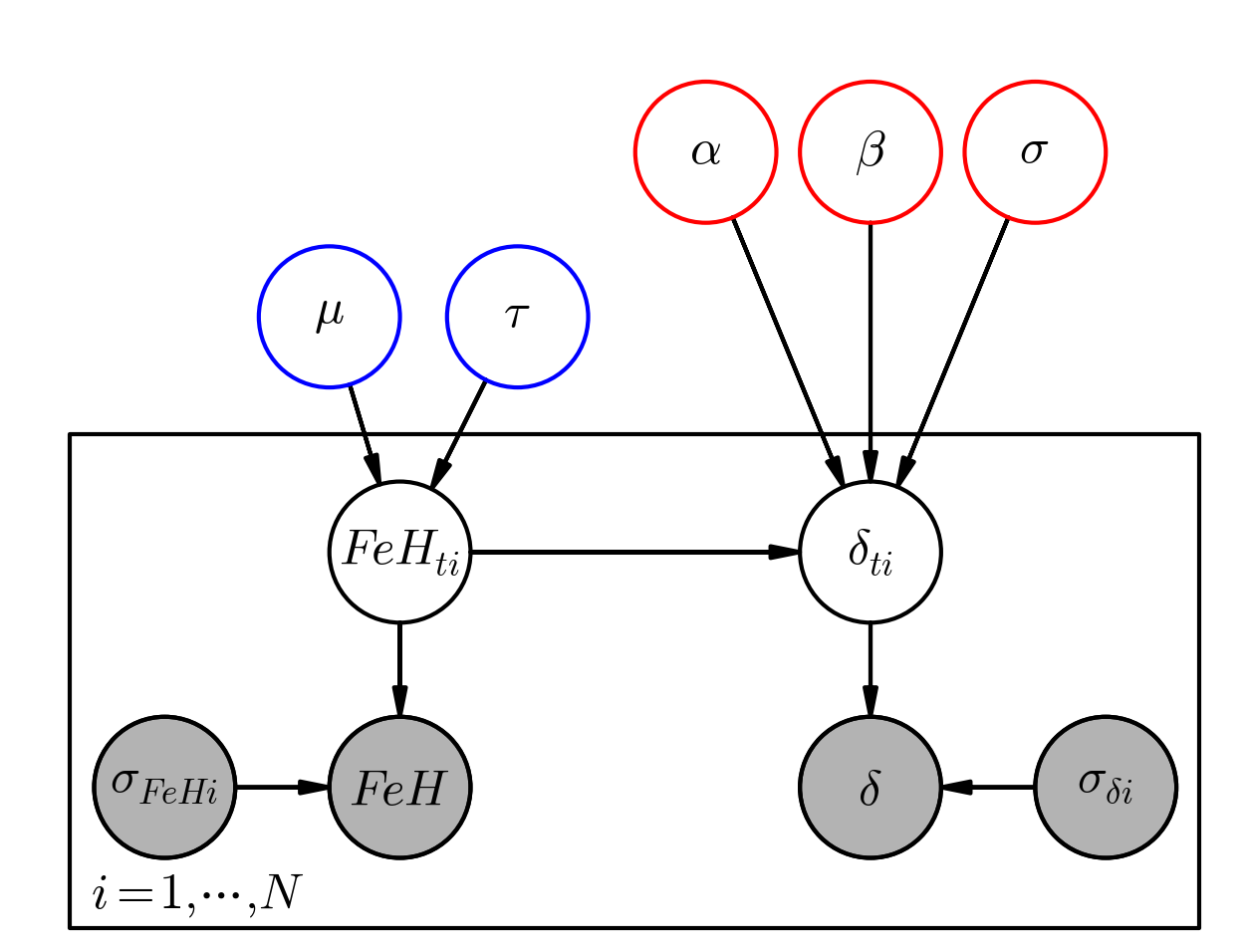}
  \caption{A graphical presentation of our HBM is given in this illustration. The graynodes are the observed parameters. The true missing parameters are in the white nodes. The blue (upper left) nodes are the nuisance parameters and the red nodes(upper right) are the parameters of interest. FeH$_i$ = stellar parameter of the i$^{th}$ planet, $\sigma_{FeH,i}$ = uncertainty on the stellar metallicity of the i$^{th}$ planet, $\delta_i$ = transit depth of the i$^{th}$ planet, $\sigma_{\delta,i}$ = uncertainties on the transit depth of the i$^{th}$ planet, FeH$_{ti}$ =  true stellar metallicity of the i$^{th}$ planet, $\delta_{ti}$ = true transit depth of the i$^{th}$ planet, $\mu$ and $\tau$ = {\it nuisance} parameters, $\alpha$,  $\beta$ and $\sigma$ = parameters of the linear model }
  \label{fig:graphical-model}
\end{figure}

Hierarchical Bayesian Modeling (hereafter HBM) allows for intrinsic scatter and heteroscedastic measurement errors i.e the uncertainties for each data points are different. We followed an approach similar to that proposed by  \cite {Kell}. We constructed the likelihood function in a simple way in order to relate the parameters of interest to the observed data, taking into account the measurements uncertainties. We extended the model to account for false positive by {\cite {Fress}}. This updated model the HBM is used to study the correlation between the transit depth ($\delta$) and the metallicity (FeH) of host stars. A graphical illustration of our HBM is given in Figure \ref{fig:graphical-model}. Markov chain Monte Carlo (hereafter MCMC) was performed using the Python package PySTAN, a package for Bayesian inference. We ran models with 4 Markov Chains, with 5000 iterations  for the first sample and 10 000 iterations for the second one. The first 50 per cent of each chain was discarded as "burn-in".
This work is the first to perform a full HBM to study correlations in general and the correlation between the transit depth and the metallicity of {\it Kepler's} giant planets, in particular. Most importantly, the quantification of the intrinsic dispersion which has not been characterized before, is now defined.

\section{Results}

\begin{figure}
\centering

{\includegraphics[scale = 0.27]{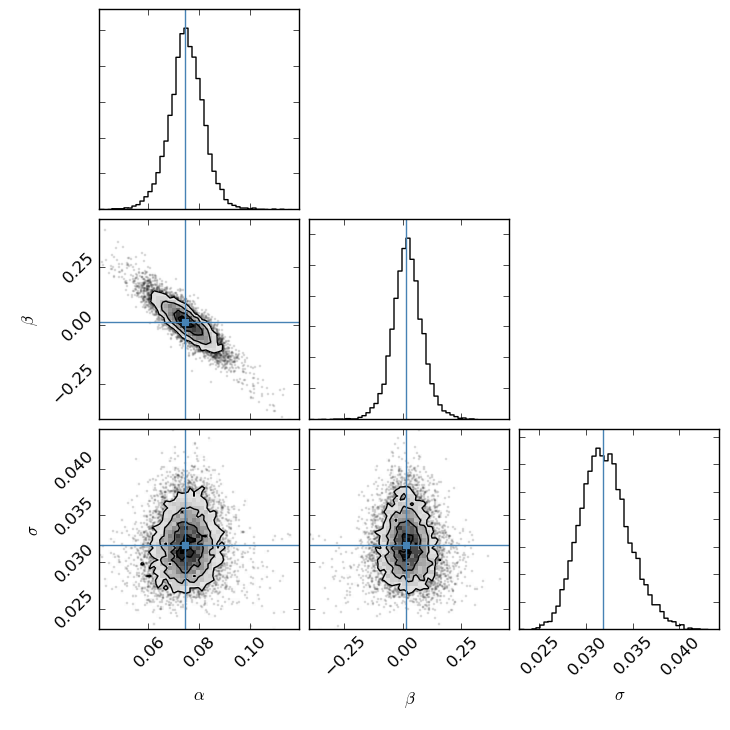}}
{\includegraphics[scale = 0.23]{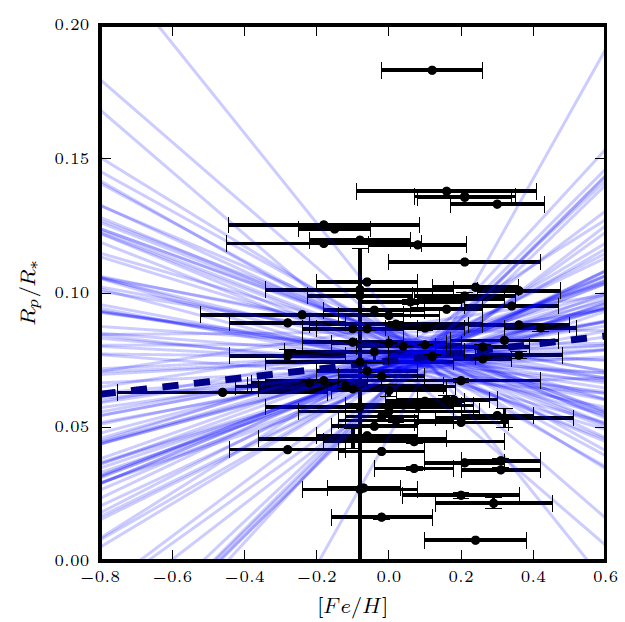}}\\

{\includegraphics[scale = 0.27]{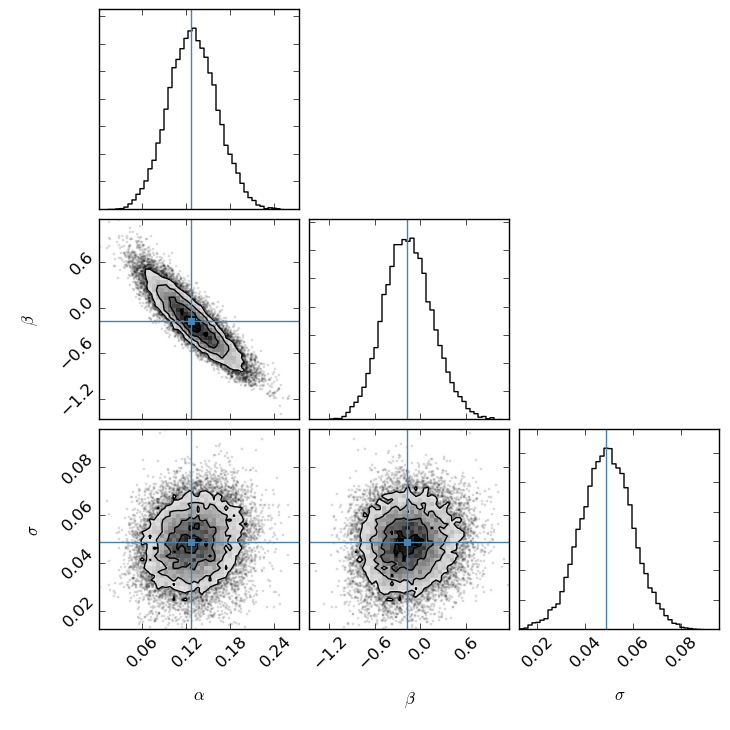}}
{\includegraphics[scale = 0.27]{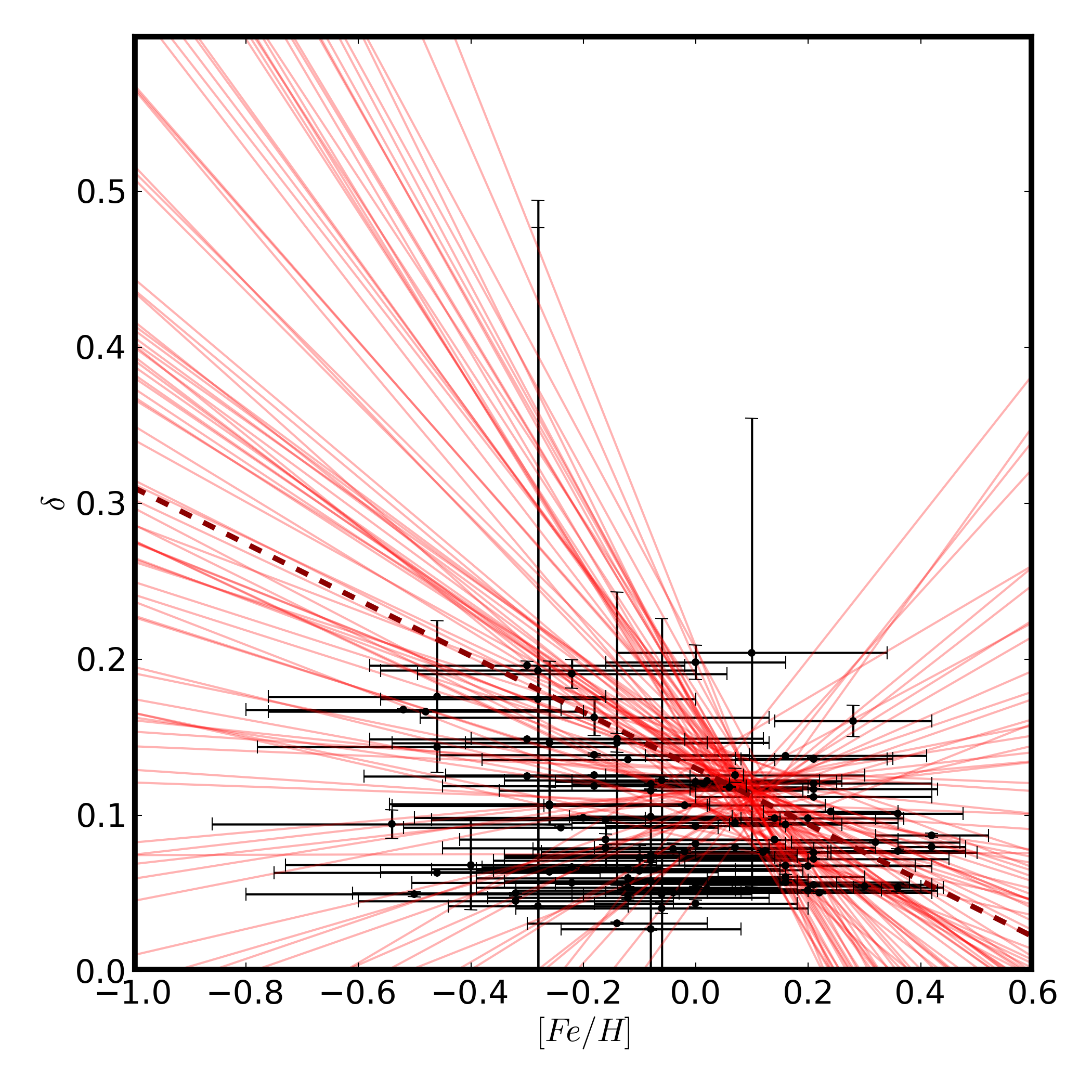}}\\

\caption{{\bf Left} column are Posterior probability distribution for the parameters of our model and as computed by the MCMC package, PySTAN, marginalized over the other parameters (left up for the confirmed sample and left down for the complete one). {\bf Right} column graphs are the transit depth ($\delta$) of {\it Kepler's} giant planets  vs. the metallicity of the host star ([Fe/H]).The dashed lines represent the best fit line and the light lines are samples from the MCMC chain.}
\label{fig:results}
\end{figure}

The posterior distributions for each of the parameters of interest ($\alpha$, $\beta$ and $\sigma$) produced by running MCMC are shown in the left panel in the Figure \ref{fig:results}, for the two samples. The equation of the "best-fit" linear models are : $\delta = (0.07\pm0.014)+(0.02\pm0.14)FeH$ with an intrinsic scatter of $\sigma = 0.03\pm0.005$ for the confirmed planets sample. For the Complet subsample the best fit and the intrinsic scatter are $\delta = (0.13\pm0.06)+(-0.18\pm0.06)FeH$ and $\sigma = 0.05\pm0.02$, respectively. The transit depth is plotted against the metallicity of the host star along with their uncertainties, in the right panel of the Figure \ref{fig:results}. The "best-fit" models (dotted lines) are shown along with 100 random samples from the MCMC chain. This shows clearly the existence of a large intrinsic scatter. A robust way to demonstrate the abscence of correlation between the transit-depth and the stellar metallicity.

\section{Discussion}

We presented for the first time in the exoplanet literature, and within a Bayesian framework, a study of the correlation between the transit depth of {\it Kepler's} giant planets and the metallicity of the host star. Data from {\it Kepler}(Q1-Q17) (\cite {MU}) allowed us to characterize the intrinsic scatter in the relation with a robust statistical analysis. We did not assume that the observed parameters have negligible uncertainties. Moreover, we did not consider all the planets in the samples as {\it bona fide}. We expand the hierarchical model presentend in \cite {Kell}, to account for the false positive rates. We also considered in the model the relevant selection effects. The use of Markov Chain Monte Carlo (MCMC) to fit models to observations is becoming a standard practice in astronomy. We performed MCMC using the package PyStan to estimate the parameters of the hierarchical linear model. We established that there is no correlation between the transit depth and the stellar metallicity of {\it Kepler's} gas giant planets. Our model indicates that there is a relatively large intrinsic scatter in the relation. Hence, the previous results could probably be an artifact which shows the importance of accounting for uncertainties and for possible false positives. This is an exciting result actually. It also proves the importance of accounting for selection effects and biases within the transit surveys, such as {\it Kepler}, and the significance of studying a complete subsample. \cite {dodo} interpreted the negative correlation as evidence that metal-rich planets of a given mass are denser than their metal-poor counterparts, leading to smaller radii (\cite {Fortnett}). On  the contrary, with our  robust statistical  model, we have proven  the independence of  transit depth and the stellar metallicity. It certainly warrants further investigations to check what planetary formation model can explain the outcome.

\end{document}